\documentclass[graybox]{svmult}

\usepackage{mathptmx}       
\usepackage{helvet}         
\usepackage{courier}        
\usepackage{type1cm}        
\usepackage{makeidx}         
\usepackage{graphicx}        
\usepackage{multicol}        
\usepackage[bottom]{footmisc}

\usepackage{cite}

\makeindex             

\begin{document}

\title*{Reactions induced by $^9$Be in a four-body continuum-discretized coupled-channels framework}
\author{J. Casal, M. Rodr\'iguez-Gallardo and J. M. Arias}
\institute{J. Casal \and M. Rodr\'iguez-Gallardo \and J. M. Arias  \at Dpto. de F\'isica At\'omica, Molecular y Nuclear, Facultad de F\'isica, Universidad de Sevilla, Apto. 1065, E-41080 Sevilla, Spain, \email{jcasal@us.es}}
%
%
\maketitle

\abstract*{ }

\abstract{We investigate the elastic scattering of $^9$Be on $^{208}$Pb at beam energies above (50 MeV) and below (40 MeV) the Coulomb barrier. The reaction is described within a four-body framework using the Continuum-Discretized Coupled-Channels (CDCC) method. The $^9$Be projectile states are generated using the analytical Transformed Harmonic Oscillator (THO) basis in hyperspherical coordinates. Our calculations confirm the importance of continuum effects at low energies. }

\section*{Introduction}
\label{sec:intro}
The $^9$Be nucleus is a stable system but presents a small binding energy below the $\alpha + \alpha + n$ threshold~\cite{Tilley04}, 1.5736 MeV. It shows also a Borromean structure, since none of the binary subsystems $\alpha + \alpha$ or $\alpha + n$ form bound states. Reactions involving $^9$Be should reflect both its weakly-bound nature and its three-body structure. Previous calculations considering $^9$Be as a two-body projectile~\cite{Pandit11} and also as a three-body projectile~\cite{Descouvemont15} show that breakup effects are important even at sub-barrier energies. 


In this work, we describe the elastic scattering of $^9$Be on $^{208}$Pb within a four-body CDCC method~\cite{Matsumoto06,MRoGa08}, considering a three-body projectile plus a structureless target. We generate the projectile states within an $\alpha + \alpha + n$ three-body model using the analytical THO basis~\cite{JCasal13,JCasal14} in hyperspherical coordinates. We pay special attention to the position of the relevant states of the system. The 3/2$^-$ ground state and the $^9$Be low-energy resonances are fixed to the experimental values. We refer the reader to Refs.~\cite{MRoGa08,JCasal15} for details about the theoretical formalism. 

\section*{Results}
\label{sec:res}
\begin{figure}[t]
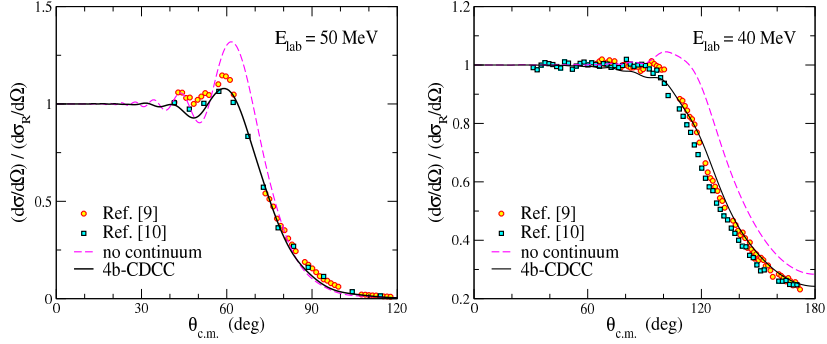

\centering
\includegraphics[scale=.3]{50.eps}\hspace{5pt} \includegraphics[scale=.3]{40.eps}
\caption{$^9$Be + $^{208}$Pb elastic cross section at 50 MeV (left panel) and 40 MeV (right panel).}
\label{fig:elas}
\end{figure}

The model space describing the $^9$Be projectile includes $j^\pi=3/2^\pm,1/2^\pm,5/2^\pm$ states. The coupled equations are solved considering the projectile-target interaction multipole couplings to all orders.  
In Fig.~\ref{fig:elas} we show the elastic cross section angular distribution in the center of mass frame, relative to the Rutherford cross section, at beam energies above (50 MeV) and below (40 MeV) the Coulomb barrier. Dashed lines correspond to calculations including the ground state only, and solid lines are the full CDCC calculations. The experimental data are from Refs.~\cite{Wolliscroft04} (circles) and~\cite{Yu10} (squares). The agreement between our calculations and the data is improved when we include the coupling to breakup channels. We confirm that this effect is important even at energies below the Coulomb barrier.

%

\begin{acknowledgement}
This work has been supported by the Spanish Ministerio de Econom\'{\i}a y Competitividad under FIS2014-53448-c2-1-P and FIS2014-51941-P, and by Junta de Andaluc\'{\i}a under group number FQM-160 and Project P11-FQM-7632.
J. Casal acknowledges a FPU grant from the Ministerio de Educaci\'on, Cultura y Deporte, AP2010-3124. M. Rodr\'iguez-Gallardo acknowledges a postdoctoral contract by the VPPI of the Universidad de Sevilla.
\end{acknowledgement}

\vspace{-10pt}

\bibliographystyle{unsrt}
\bibliography{./bibfile}
\end{document}